# Prediction of tubular solar still performance by machine learning integrated with Bayesian optimization algorithm

Yunpeng Wang [a, #], A. W. Kandeal [a, b, #], Ahmed Swidan [c], Swellam W. Sharshir [a, b]*, Gamal B. Abdelaziz [d], M. A. Halim [d], A.E. Kabeel [e], Nuo Yang [a] *

[a] State Key Laboratory of Coal Combustion, Huazhong University of Science and Technology, Wuhan 430074, China.
[b] Mechanical Engineering Department, Faculty of Engineering, Kafrelsheikh University, Kafrelsheikh 33516, Egypt.
[c] Mechanical Department, Faculty of Industrial Education, Helwan University, Cairo, Egypt
[d] Mechanical Department, Faculty of Industrial Education, Suez University, Suez, Egypt
[e] Mechanical Power Engineering Department, Faculty of Engineering, Tanta University, Tanta, Egypt.

[#] equally contributed on this work.

*Corresponding authors: SWS (swellamali@yahoo.com); NY (nuo@hust.edu.cn)

# Abstract

**Presented is a new generation prediction model of a tubular solar still (TSS) productivity utilizing two machine learning (ML) techniques, namely: Random forest (RF) and Artificial neural network (ANN).** Prediction models were conducted based on experimental data recorded under Egyptian climate. Meteorological and operational thermal parameters were utilized as input layers. Moreover, Bayesian optimization algorithm (BOA) was used to obtain the optimal performance of RF and ANN models. In addition, these models' results were compared to those of a multilinear regression (MLR) model. As resulted, experimentally, the average value accumulated productivity was 4.3 L/(m$^2$day). For models' results, RF was less sensitive to hyper parameters than ANN as ANN performance could be significantly improved by BOA more than RF. In addition, RF achieved better prediction performance of TSS on the current dataset. The determination coefficients ($R^2$) of RF and ANN were 0.9964 and 0.9977, respectively, which were much higher than MLR models, 0.9431. Based on the robustness performance and high accuracy, RF is recommended as a stable method for predicting the productivity of TSS.

# 1. Introduction

**Currently, despite 75 % on the globe is bodies of water, shortage in drinkable water is a catastrophic issue that the world faces proportionally to increase in population, industrial activities, and agricultural progress; hence, 800 million people lack potable water access**. Also, unfortunately, the available amount of drinkable water is very small as 97% plant's water resources are salty and the capital amount of freshwater is in glaciers, icecaps, and underground. In addition, about half of global water is expected to be consumed by 2050 [1]. Along with that, researchers have thought to positively face this problem via many technologies of water desalination. Many effective methods of desalination have been investigated like multistage flashing [2, 3], Reverse osmosis [4, 5], and multi-effect distillation and vapor compression [6, 7]. Despite being effective and capacitive technologies on a large production scale, these methods need high-power sources like fuel and electricity, high installation and maintenance costs, and highly experienced operators and workers. Solar desalination is mostly off the aforementioned demerits and hence, it has become promising technology especially for arid and semi-arid areas, and low to moderate demand capacity.

**Solar still (SS) is the most common solar desalination system that features for its simple construction, low installation and maintenance cost, as well as long-life operation. However, undesirably, it achieves low productivity. Hence, many modifications have been done.** Conventional SS consists of an inclined glass cover, a rectangular basin that contains saltwater, and thermally insulated from both sides and bottom. Some modifications have been conducted on basin; such as: stepped SS [8], finned SS [9] and wick SS [10]. Some other modifications have been done on the glass cover; such as: double slope SS [11], and pyramid SS integrated with nanofluid and evacuated tubes [12].

**Recently, tubular solar still (TSS) is one of the most efficient SS systems. It mainly consists of a tubular transparent cover and water basin. The tubular cover helps the basin to be ever exposed to solar radiation from any direction eliminating shadow effect in contrary to other designs that receive solar radiation only from upward causing shadow effect from sidewalls.** Furthermore, due to rounded shape of the cover, all condensed water can be easily collected compared to other designs in which condensation occurs on inclined flat surfaces. Many investigations have been conducted to improve the performance of TSS [13], such as: controlling the basin water depth and cover cooling [14]; providing parabolic concentrator solar tracking system besides surface cooling of either whole surface or gap between two concentric tubes [15]; and using different shapes of absorber such as semi-circular corrugated one [16].

**For extending the prediction of SSs' performance, many models have been proposed which can be categorized into three methodologies.** First methodology depends on the numerical solution of differential equations of heat and mass transfer [17, 18]. Second one is a regression model that can predict the relationship between multi-factor inputs and output of the system. [19, 20]. The third method, current work topic, utilizes one of machine learning (ML) and artificial intelligence technologies which have been commonly used in many scientific aspects particularly energy systems because of being able to access the key information of an energy system based on real experimental data. ML methods are considered to be more preferable than either simulation models or experimental work.

**Artificial neural network (ANN) are a well-known machine learning method that is inspired by the human brain to learn the complex nonlinearity relationships in specific problems [21]. It has been successfully applied in various engineering systems [22-24].**

**Many ANN models have been conducted to give a complete prediction of desalination systems.** Abujazar et al. [1] presented a cascaded forward neural network model to predict the productivity of inclined stepped SS based on experimentally recorded data and compared with data resulted from regression and linear models. The model was more accurate prediction than either regression or linear models [25]. Also, productivity($m_d$), operational recovery ratio (ORR), and thermal efficiency ($\eta_{th}$) of a SS was predicted using ANN model by Mashaly et al. [26]. The system proved that the model was accurately valid for prediction with the determination coefficient ranged from 0.94 to 0.99 for predicted parameters. In addition, instantaneous thermal efficiency of inclined passive SS fed with agricultural drainage water was forecasted using ANN model developed by Mashaly and Alazba [27]. As resulted, the most appropriate model was that had six neurons and a hyperbolic tangent transfer function and optimal model have 7–6–1 architecture with 0.949 mean coefficient of determination. The freshwater productivity was further predicted [28] by using ANN model at which five neurons; from which, hyperbolic tangent transfer function was the most appropriate, and the optimal model had 7–5–1 architecture with coefficient of determination of 0.96. Moreover, ANN proved its ability to predict the $\eta_{th}$ of triple basin SS according to a study conducted by Hamdan et al. [25] in which three models were used namely: Feedforward, Elman, and Nonlinear Autoregressive Exogenous (NARX) networks. As resulted, the feedforward model achieved the best performance estimation while NARX was the worst. On the other hand, real operational experimental data is not only the input for ANN predictions of SS productivity, but daily weather observations can be used as well, as confirmed by Santos et al. [29].

Random forest (RF) [30] is a powerful supervised machine learning algorithm inspired by ensemble technique. RF typically works by combining numbers of decision trees into one algorithm model to improve performance. It is applicable in various aspects, such as statistics [31-33], materials [34-37] and biology[38, 39].

**Based on the literature review, despite being an effective way in SSs' performance modeling,** the numerical methods require highly specified computers with complex modeling systems besides neglecting some correlative factors and utilizing ideal hypotheses and hence, accuracy and validity of the model are affected. On the other hand, regression model is inaccurately used in predicting the SS systems performance [40]. Due to the necessity for accurate and reliable modeling of the productivity of SS systems on an hourly basis, ML can be an alternative way of dealing with complex problems [41].

**This work aims to predict hourly productivity of TSS using two ML methods; namely: such as RF and ANN, besides optimizing them by** BOA**.** First, a series of experiments were conducted on the test rig in 16 days, 9 hours each (144 hours in total). Second, productivity prediction (output) was conducted using two ML methods (RF and ANN) based on other measured data (input). Then, to obtain the best performance optimal performance of RF and ANN models, BOA was used for adjusting hyper parameters of models. These ML methods built in this work are constructed on Python packages Scikit-learn [42], Panda [43] and Numpy [44]. Different data sizes were used in model training. Furthermore, the results were compared to multilinear regression (MLR) model. Both experimental and prediction results are presented as well as feature importance of input data variables.

## 2. System description and instrumentations

**The proposed TSS system, shown in Fig. 1 and Fig. 2, consists of a tubular cover, basin, feedwater assembly, and required instrumentation for complete parameters' measurements.** A 1.5-thick polycarbonate transparent cylinder of 50 cm diameter and 100 cm

length was used as tubular cover permitting all direction exposure of basin to solar radiation. In addition, the basin was made of black-painted steel of 90, 40, 5 and 0.15 cm for length, width, side-wall height, and thickness, respectively. Inside the basin, saline water height was limited at 0.5 cm through a feeding system that supplied feed water as same as the hourly-evaporated amount. This height was primarily set through a graded level-meter attached to every side of the basin.

**The system was well instrumented to measure the performance parameters which were hourly recorded**: solar radiation intensity ($I_R$), wind speed ($V_w$), hourly productivity ($m_h$), accumulated productivity ($m_d$), and temperatures of basin plate ($T_b$), saltwater ($T_{sw}$), glass cover ($T_{gc}$), and ambient ($T_\infty$). The freshwater productivity is collated in an 8-liter calibrated bottle with an accuracy of ±8 ml. Also, Thermocouples of K- type with a range of (- 50 to 280 °C) was used for temperature measurements. All temperatures were monitored through a digital thermometer having an accuracy of ± 0.1 °C. In addition, $V_w$ was measured by a fan-type anemometer with (0 to 45 m/s) range and (±1 m/s) accuracy. Moreover, $I_R$ was measured by TES-1333 solar meter of (0 to 5000 W/m$^2$) range and (±5 W/m$^2$) accuracy.

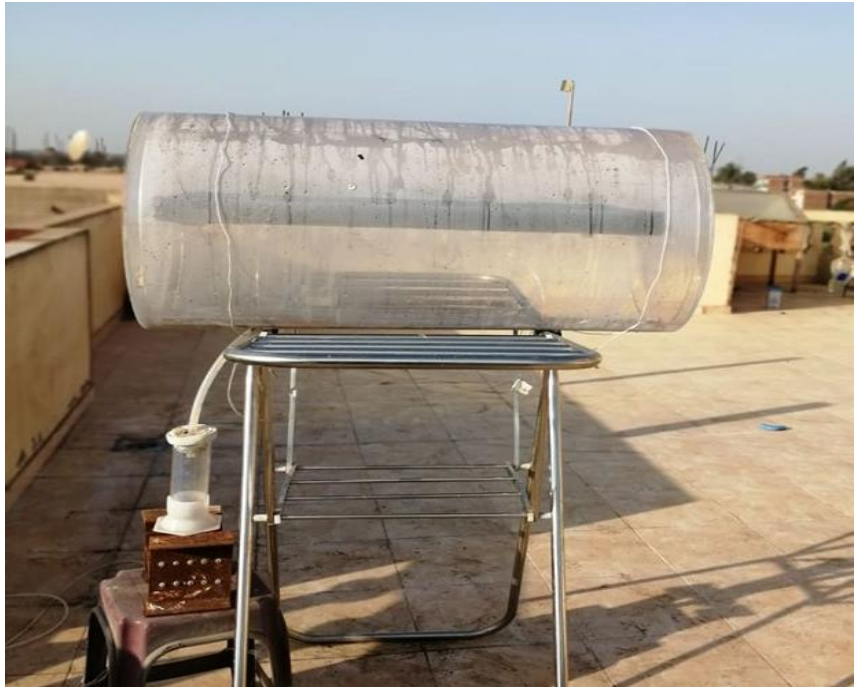

Fig. 1 Photo of TSS experimental set up.

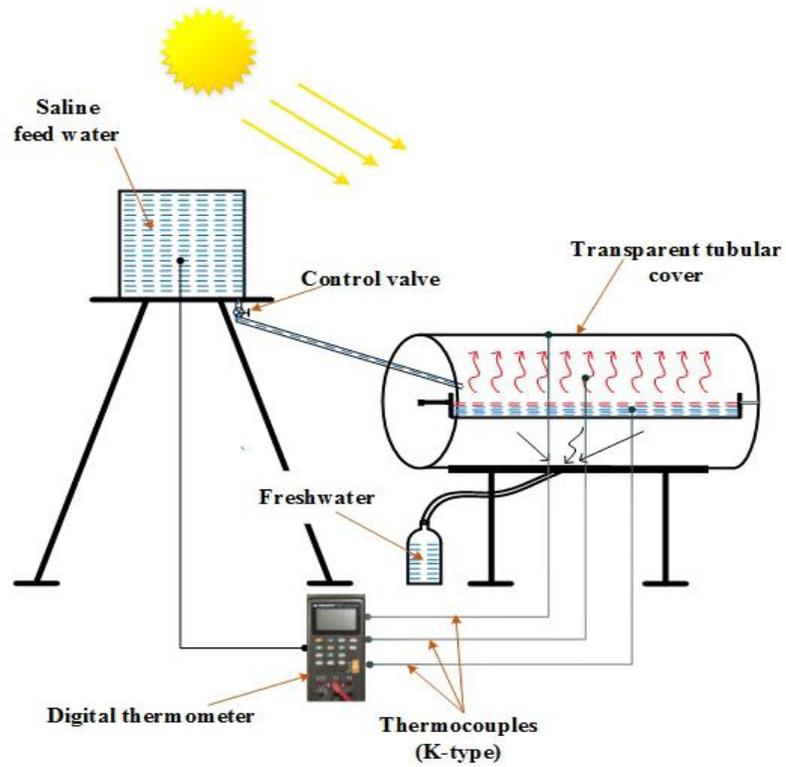

Fig. 2 Schematic diagram of TSS experimental set up.

## 3. Methodology of machine learning

For a given TSS dataset, $M = \{X,y\}_{1:t}$, y is the objective value (hourly production) corresponding to input parameter $X$ (for example air speed, air temperature, etc.), and t is the number of data samples. When there are $N$ input parameters, $M \in R^N$. Then, the details of establishing ML model can be summarized as follows:

### 3.1 Random forest

RF retains many benefits of decision trees while achieving better results through the ensemble of decision trees. The final results of RF are averaged over the decision trees to decrease the variance and increase the accuracy. In current work, RF is combined with classification and regression trees (CART) [35], which have proved powerful results in many fields. The main process shown in Fig. 3 can be concluded into three steps.

**(1) Data manipulation**

First, the bootstrap resampling method is performed for randomly generating *ntree* sets of data from dataset M. According to the specific proportion, each set of data is split to a training set and testing set. In order to decrease the variance between different features and improve the convergence rate of the model, all sets of data are normalized to mean 0 and standard deviation 1 in Scikit-learn package [42].

**(2) Model construction**

After data manipulation, *ntree* decision trees will be grown on *ntree* sets of data. Based on the particular random algorithm in tree construction, each decision tree is different from others which will ensure the diversity and ensemble performance of the final forest. The final result of RF is averaged by K decision trees to decrease the variance and increase the accuracy.

**(3) Model optimization**

In order to ensure the trade-off between data overfitting and high accuracy of the RF model, Bayesian optimization is conducted for adjusting hyper-parameters automatically. The objective function is the accuracy of five-fold cross-validation on the training set.

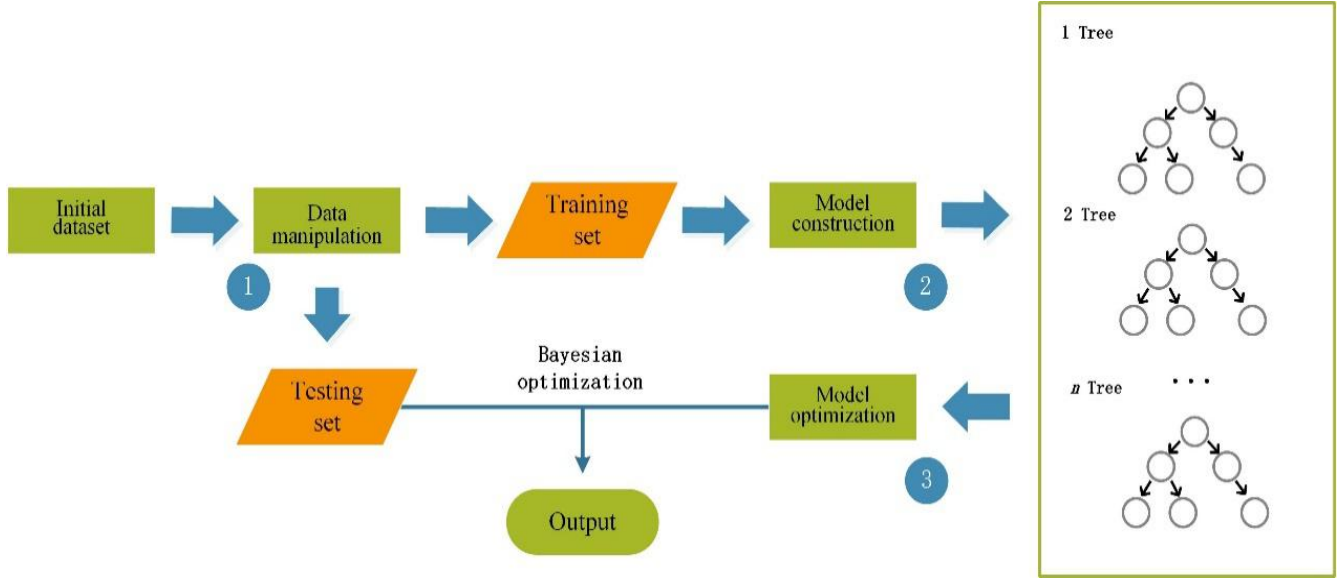

Fig. 3 Data flow diagram of RF.

After model validation, RF can measure the importance of certain features which is defined as summation of Gini index (impurity) reduction overall nodes by using this feature[36, 45]. It can be calculated as follows:

$$\text{IM}_i = \sum_j G_{i-\text{before}}(j) - G_{i-\text{after}}(j) \qquad (1)$$

where $\text{IM}_i$ is the importance of feature i, $G_{i-\text{before}}(j)$ is the Gini index (impurity) of node j before node splitting by descriptor i, while $G_{i-\text{after}}(j)$ is the Gini index (impurity) after node splitting. The Gini index $G(j)$ is defined as:

$$G(j) = 1 - \sum_{a \in A} p_a^2 \qquad (2)$$

where G(t) is the Gini index (impurity) of node t, $p_a$ is the relative frequency of class a in the node j.

## 3.2 Artificial neural network

Different from other ML algorithms, ANN makes no prior assumptions about the data distribution and can highly model non-linear functions.

As shown in Fig. 4, the typical ANN usually consists of three parts, input layer, hidden layer and output layer. The nodes in the input layer equal the number of features in input data and the output layer represent the objective value of the dataset. Based on the complex connections between neurons of the hidden layer and the activation function in each node, ANN can build a nonlinear mapping between input and output. The main steps of training ANN can be expressed as follows:

1) Based on the dataset M, split training set and testing set according to the specific proportion. Subsequently, initialize ANN weights to a small random value and propagate training set data through the network to obtain the prediction value of hourly production. For one training sample $(x_k, y_k)$, the error E can be calculated by equation (3):

$$E_k = \frac{1}{2}(y_0 - y_k)^2 \tag{3}$$

where $y_0$ is the value of the output layer.

2) Assume the weights between hidden layer and output $y_0$ is $w_{iy}$, the weights between the input layer and the hidden layer is $v_{xj}$, i,j are the number of neurol in hidden layer. Based on backpropagation algorithms, propagate the error signal back through the network and adjust the weights between nodes, the $w_{iy}$ can be updated in equation (4):

$$w_{iy\_new} = -\theta \frac{\partial E_k}{\partial w_{iy}} + w_{iy} \qquad (4)$$

where

$$\frac{\partial E_k}{\partial w_{iy}} = \frac{\partial E_k}{\partial y_o} \cdot \frac{\partial y_o}{\partial \beta_i}$$

$\beta_i$ is the input of output layer.

Similarly, $v_{xj}$ can be updated in equation

$$v_{xj\_new} = -\theta \frac{\partial E_k}{\partial b_j} \cdot \frac{\partial b_j}{\partial \alpha_j} + v_{xj} \qquad (5)$$

where $b_j$ is the output of jth neurol in hidden layer, $\alpha_j$ is the input of jth neurol in hidden layer. $\theta \epsilon (0,1)$ is the learning rate of ANN. Large $\theta$ means the fast convergence rate but may lost in Local optimum, while the small $\theta$ means the accuracy and slow convergence rate.

3) Repeat steps 1-2 until the terminal condition is reached.

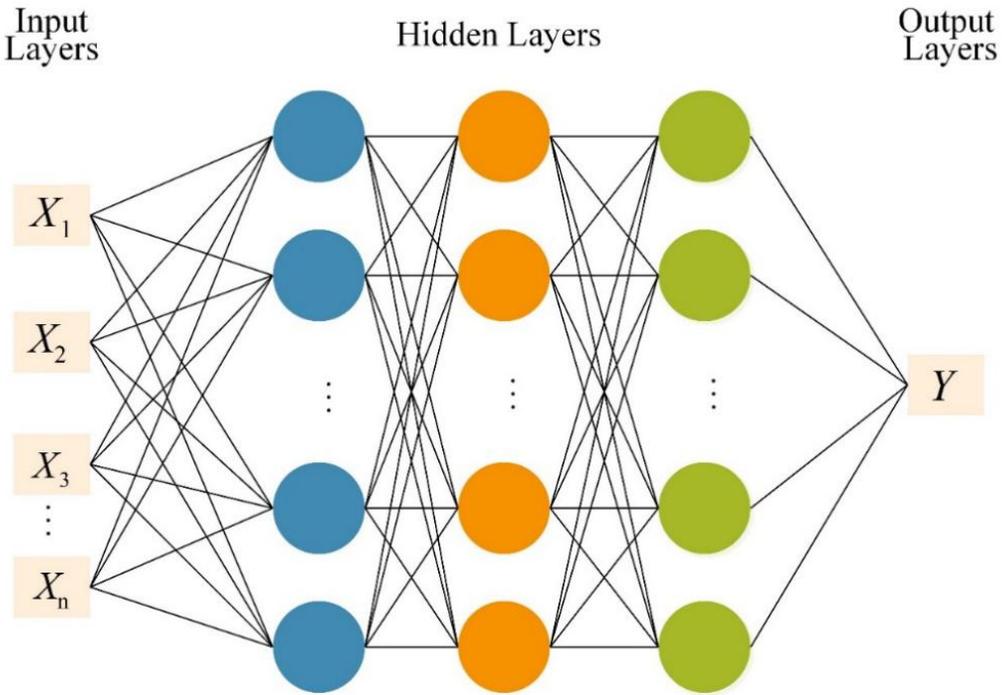

Fig. 4 Data flow diagram of ANN.

## 3.3 Bayesian optimization

In order to obtain global optimal performance of RF and ANN model, the Bayesian optimization algorithm (BOA) is conducted for searching the most appropriate hyper-parameters in the current work. BOA is a powerful technique for finding the extrema value of black-box function and it's particularly useful when evaluations of the objective function are costly[46, 47]. The typical BOA[48] works by using the known observations to fit a Gaussian process model and get the posterior sample location by the Gaussian process model.

In the current study, for the sake of avoiding the risk of overfitting, the result of 5-fold cross-validation is set as the objective function of BOA. Assume $X = a_1, a_2 ... a_i$ are hyper-parameters of RF or ANN, Dataset $D = \{X_{1:n}, Y_{1:n}\}$ are samples observed in the previous test. $X$ means one potential set hyper-parameters of RF or ANN; $Y$ represent the result of 5-fold cross-validation corresponding to $X$. The procedure for BOA is as follows:

First, the Gaussian process model M is established on accumulated observations D in equation (6)

$$M \sim N(0, K) \tag{6}$$

where kernel matrix K is

$$K = \begin{bmatrix} k(x_1, x_1), \cdots k(x_1, x_t) \\ \vdots \qquad \vdots \\ k(x_t, x_1) \cdots k(x_t, x_t) \end{bmatrix} \tag{7}$$

$$k(x_i, x_j) = \exp\left(-\frac{1}{2} \|x_i - x_j\|^2\right) \tag{8}$$

Second, the next location $x_{t+1}$ to sample, according to acquisition function, is determined at which the observation property M is expected to be the best (the high accuracy of 5-fold cross-validation). Subsequently, the gaussian process model M is updated by including the new observation $x_{t+1}$.

The two steps are repeated until the terminal condition is reached.

### 3.4 Evaluation indicators

The predictive performance of the model is evaluated by four statistical criteria; namely: mean absolute error (MAE), mean absolute percentage error (MAPE), mean square error (MSE) and determination coefficient ($R^2$). The formula is shown in Table 1, the smaller MAE, MAPE, MSE and the closer of $R^2$ value to 1 mean the better performance.

Table 1 Expression of evaluation indicators.

| Evaluation indicators | Expression |
|---|---|
| Mean absolute error (MAE) | $\text{MAE} = \frac{1}{n}\sum_{i=1}^{n}|f_i - y_i|$ |
| Mean absolute percentage error (MAPE) | $\text{MAPE} = \frac{1}{n}\sum_{i=1}^{n}\left|\frac{f_i - y_i}{y_i}\right| \times 100\%$ |
| Mean square error (MSE) | $\text{MSE} = \frac{1}{n}\sum_{i=1}^{n}(f_i - y_i)^2$ |
| Determination coefficient ($R^2$) | $R^2 = 1 - \frac{\sum_{1}^{n}(y_i - f_i)^2}{\sum_{1}^{n}(y_i - \bar{y})^2}$ |

## 4. Result and discussion

### 4.1 Experimental results

The prediction model was conducted based on a set of experimental results recorded in 16 days. The measured parameters were $m_h$, $m_d$, $I_R$, $V_w$, and temperatures at different locations ($T_b$, $T_{sw}$, $T_{gc}$ and $T_\infty$). The average values of recorded variables during experiment days are shown in Fig. 5. According to Fig. 5 (a), average $I_R$ ordinarily increased from 554 W/m² till peak value of 1031 W/m² at noon and then decreased to 451.25 W/m²; and hence, all various temperatures followed the same profile but the hour of peak value differed. That to say, due to the heat capacity of the basin, water and glass cover, their average temperature reached a maximum value at 13:00 i.e. still heated even with a decrease of solar radiation intensity reaching maximum values of 68.8, 67.8, 53.3°C. Moreover, due to small utilized water thickness, values of $T_{sw}$ closely approached that of basin with only 1°C difference. Also, ambient and feedwater had near temperature values with differences ranged from 2.3 to 5.3°C for the sake of ambient air.

Furthermore, Fig. 5(b) presents the variation of both average $m_h$ and $m_d$. As resulted, the maximum obtained value of $m_h$ was 0.78 L at the 11$^{th}$ day; whilst its maximum average value was 0.73 L. In addition, the maximum gained value of $m_d$ was 4.7 L/day; whilst its average value was 4.3 L/day.

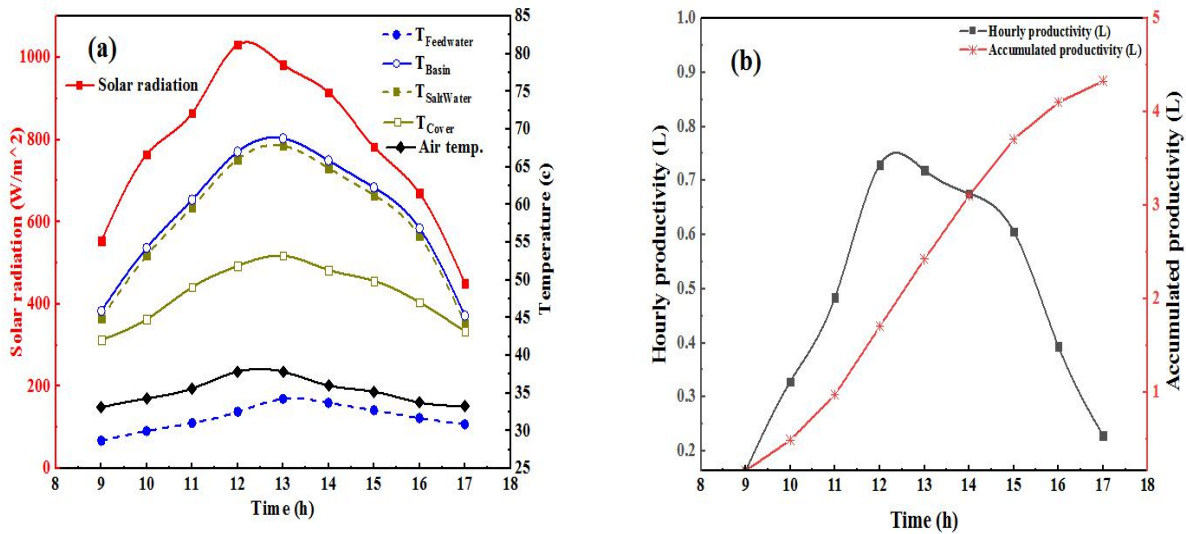

Fig. 5 Variation of average values of experimental data vs. time: (a) solar radiation and different temperatures, (b) hourly and accumulative productivity.

### 4.2 Prediction results of machine learning models

Fig. 6 displays comparisons of the prediction performance constructed by different ML models. The initial test size was set to 0.2. As shown in Fig. 6, the RF and ANN model optimized by BOA could get a higher $R^2$ closer to 1, which means a good prediction between simulation results and target values. Moreover, the similar performance between the training set and test set indicated that BOA-RF and BOA-ANN both had good generalization performance. Hence, the powerful model can be applied to the prediction of TSS with a small expense on measurement. However, the results of ANN still exist some discrepancies between RF and BOA-ANN which mean more errors have been caused in ANN prediction. It also can be seen that the difference between the result of RF and BOA-RF was not much difference. It reveals that ANN was more sensitive to hyper-parameters than RF and BOA could perform well in finding the extrema of ML model with black-box function.

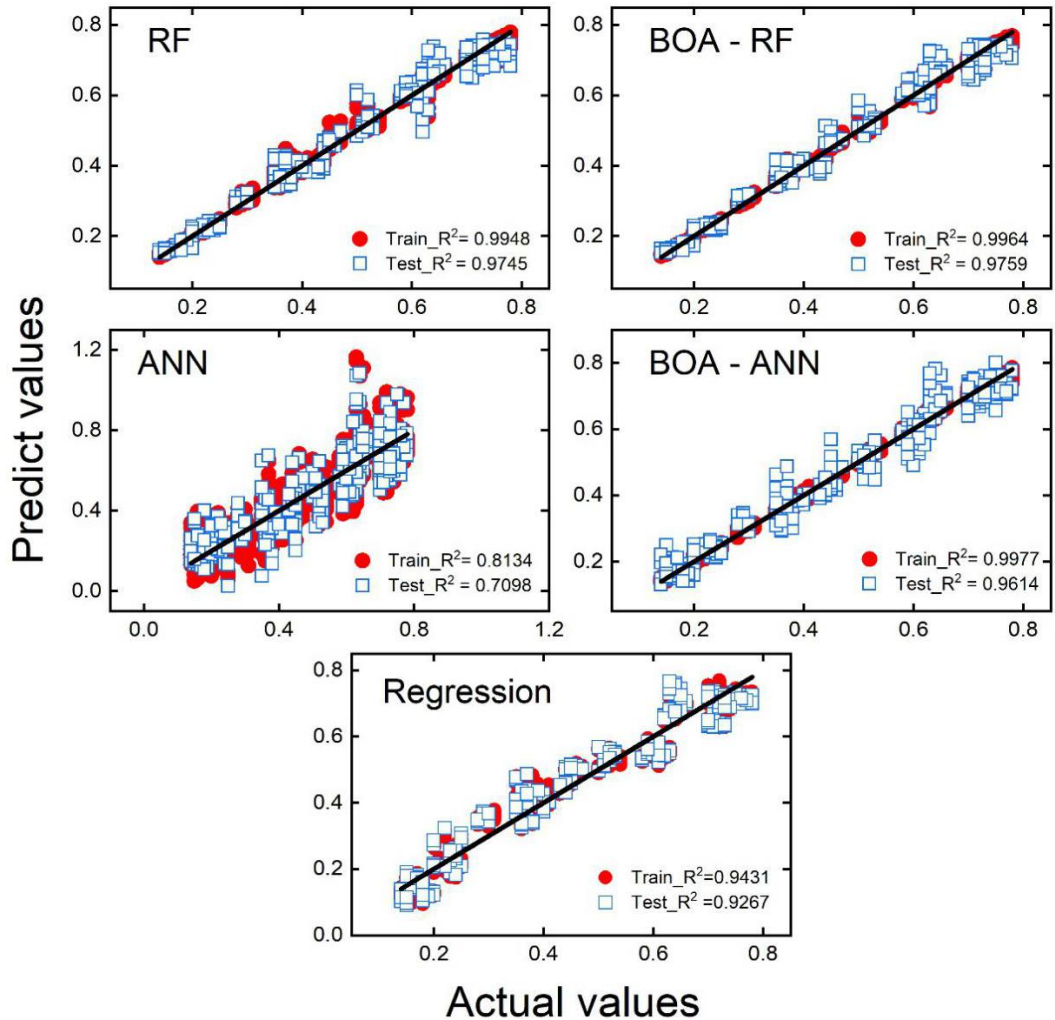

Fig. 6 Prediction results of different models

The above results can be also observed from the absolute percentage error shown in Fig. 7. In both training sets and test sets, the errors of RF did not change much with the optimization by BOA. Whereas the performance of ANN represented huge improvement after the adjustment of hyper-parameters by BOA. In addition, the BOA-RF model could achieve little better performance than BOA-ANN model.

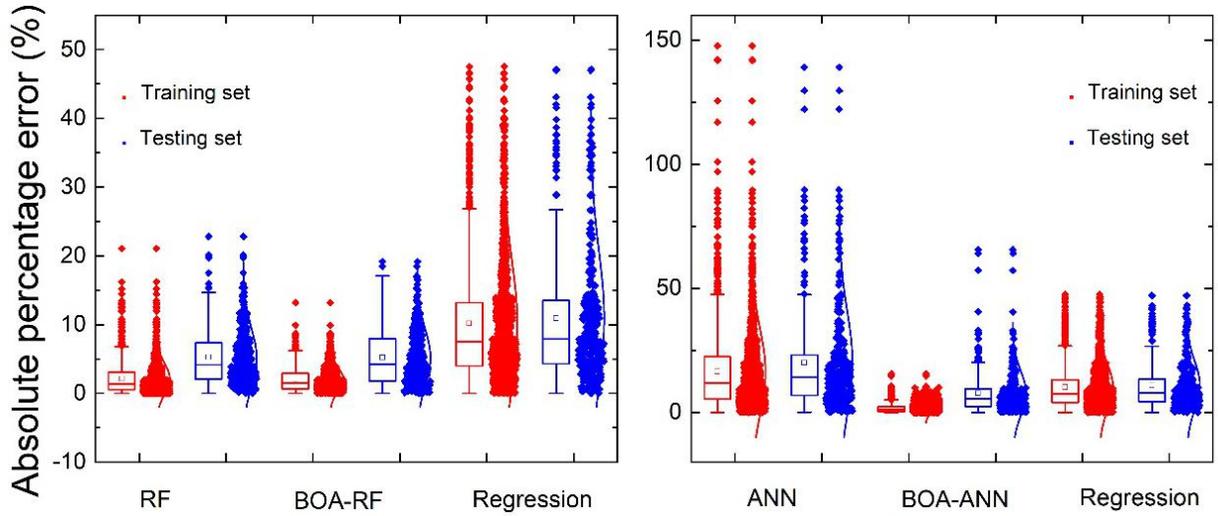

Fig. 7 Absolute percentage error in different model

Table 2 Prediction performance of evaluation indicators

| Models | | Evaluation indicators | | | |
|---|---|---|---|---|---|
| | | MAE | MAPE (%) | MSE | $R^2$ |
| RF | a | 0.0094 | 2.119 | 0.00021 | 0.9948 |
| | b | 0.0239 | 5.2372 | 0.00104 | 0.9745 |
| ANN | a | 0.0662 | 16.434 | 0.00766 | 0.8133 |
| | b | 0.08307 | 19.974 | 0.0119 | 0.7098 |
| BOA-RF | a | 0.0088 | 1.991 | 0.00014 | 0.9964 |
| | b | 0.0234 | 5.210 | 0.00094 | 0.9758 |
| BOA-ANN | a | 0.0067 | 1.6814 | 0.00009 | 0.9977 |
| | b | 0.0300 | 7.697 | 0.00157 | 0.9614 |
| MLR | a | 0.0404 | 10.166 | 0.00236 | 0.9431 |
| | b | 0.0438 | 10.910 | 0.00278 | 0.9267 |

(a is the performance of train dataset and b is for test dataset)

To further quantify the performance between different models, Table 2 shows the results of evaluation indicators in different models. Besides $R^2$ value discussed above, the results of MAE, MAPE and RMSE also had a similar trend with $R^2$ in a different model. Moreover, regardless of the training set and testing set, there existed no great variation in the MAE, MAPE, RMSE. It can be concluded that our model avoided the risk of overfitting and obtained high accuracy as well as powerful generalization performance.

### 4.3 Effect of the size of dataset

Considering the limited size of the initial dataset, to further research the influence of the dataset on predict performance, Fig. 9 shows the results of different test sizes in train/test spilled. As can be seen, with the increase of training set (decrease of the test size), there is no significant difference between different models. This indicates that the current dataset was enough to get a convergent solution.

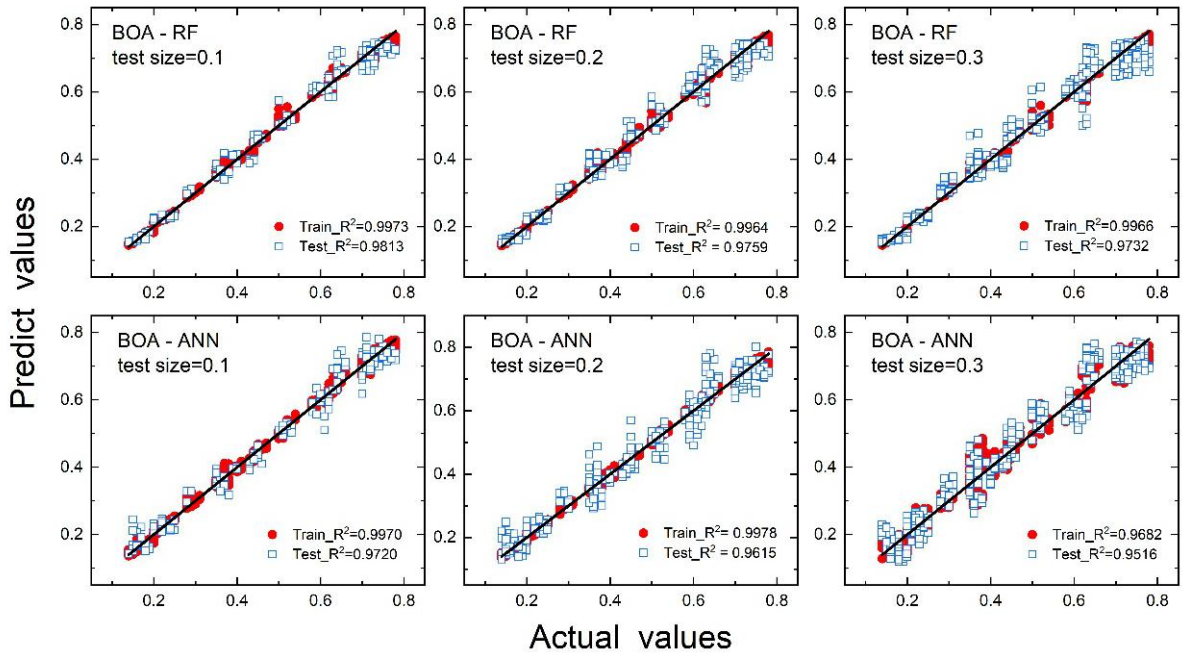

Fig. 9 Prediction results of different test size

### 4.4 Prediction results of multilinear regression model

The regression model is one of the most common statistic models for predicting the relationship between multi-factors input and the output of the system. In this study, the multiple linear regression (MLR) was conducted to predict the hourly productivity of the TSS system based on the same input data of other ML models. The model equation is shown in Eq. (7) and the detail of evaluation indicators is shown in Table 2. The determination coefficient ($R^2$) on the training set and testing set were 0.9431, 0.9267, respectively. As compared with RF and ANN in Fig.6 and Fig.7, the prediction performance of the MLR model was worse than ML methods.

$$P = 0.486 + 0.0465T + 0.0226T_\infty + 0.125I_R + 0.00249V_w + 0.0143T_w + 0.041T_b - 0.0218T_{sw} + 0.0296T_{gc} \tag{7}$$

### 4.5 Feature importance

As mentioned above, RF can calculate feature importance by intrinsic property. According to the results shown in Fig. 10, the order of feature important of parameters are in a logic order from direct to less effect on the productivity. The most important parameters are $T_{sw}$ and $T_b$ by about 40.87 and 32.43 % because they directly affect the water evaporation rate which is the major parameter of productivity. $T_{sw}$ has higher feature importance than $T_b$ as the evaporation occurs at the water surface. Subsequently, $I_R$ have effectiveness importance by about 18.2% as it, indirectly, affects the production rate. In addition, $V_w$, $T_{gc}$ and $T_\infty$ have approximately the same importance in the range of about 1.9 to 2.6% as they do not participate in the physical processes of desalination (evaporation and condensation) and they represent the small amount of heat losses from the system to surrounding. Furthermore, putting the days in either arranged or random orders has insignificant importance by 1.6 %.

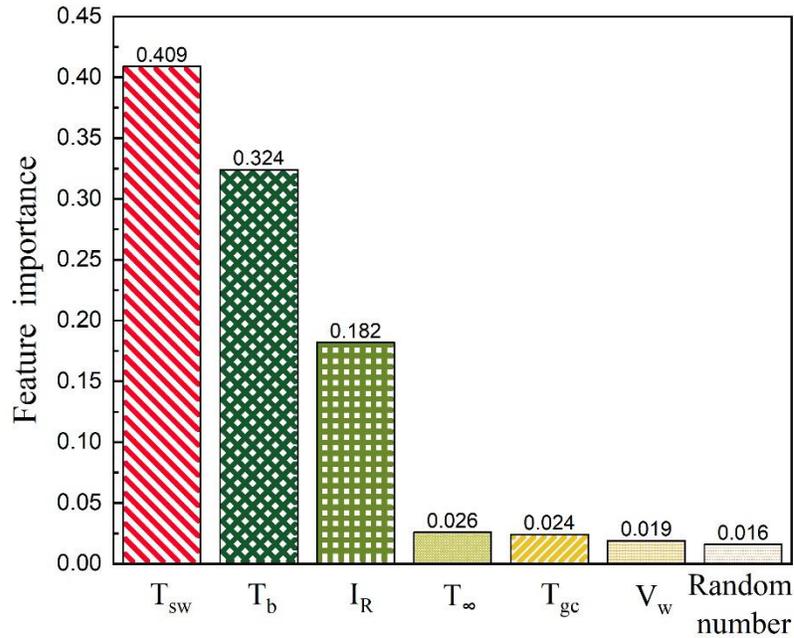

Fig. 10 Results of feature importance.

## 5. Conclusions

In this study, two prediction machine learning models of tubular solar still (TSS) were established based on random forest (RF) and artificial neural network (ANN). The Bayesian optimization algorithm (BOA) was conducted to optimize the hyper parameters of RF and ANN. These models' results were compared to those of multilinear regression (MLR) model. Models were conducted based on experimental data recorded in 16 days with 8 hours each (144 hours in total). The following results can be concluded:

- Experimentally, the maximum values of hourly and accumulated productivity were 0.78 L and 4.7 L/day, respectively.

- The prediction performance of RF without BOA on test dataset was: MAE = 0.0239, MAPE = 5.237 %, MSE = 0.00104, $R^2$ = 0.9745; while the BOA-RF was: MAE = 0.0234, MAPE = 5.21 %, MSE = 0.00094, $R^2$ = 0.9758.
- The prediction performance of ANN without BOA on test dataset was: MAE = 0.0831, MAPE = 19.974 %, MSE = 0.0119, $R^2$ = 0.7098; while the BOA-ANN was: MAE = 0.03, MAPE = 7.697 %, MSE = 0.00157, $R^2$ = 0.9614.
- The prediction performance of MLR on test dataset was: MAE = 0.0438, MAPE = 10.911 %, MSE = 0.00278, $R^2$ = 0.9267.
- The parameters that had the highest feature importance were: saltwater temperature, basin temperature, and solar radiation, of 40.87, 32.43, and 18.2%, respectively.

Thus, the results demonstrated that it is possible to predict hourly production closer to true experimental observations using ML techniques. Moreover, the performance of ANN can be significantly improved by BOA rather than RF. This indicates that RF is less sensitive to hyper parameters than ANN. In other words, RF is a more robust model than ANN. Although ANN has well predicted on many large datasets, RF achieved better prediction performance of TSS on the current dataset. Based on the robustness performance and high accuracy, RF is recommended as a stable method for predicting the productivity of TSS.

**Conflicts of interest**

There are no conflicts of interest to declare.

**Acknowledgment**

N.Y. was sponsored by National Natural Science Foundation of China (No. 51576076 and No. 51711540031), Natural Science Foundation of Hubei Province (2017CFA046) and Fundamental Research Funds for the Central Universities (2019kfyRCPY045). S. W. S. was sponsored by National Natural Science Foundation of China (No. 51950410592). The authors thank the National Supercomputing Center in Tianjin (NSCC-TJ) and China Scientific Computing Grid (ScGrid) for providing assistance in computations.